\documentclass[paper]{JHEP3} 
\usepackage{bm} % use as \bm{} to make Greek lett. bold. \boldsymbol is as good as \bm as well.

\title{The Integrability of Pauli System in Lorentz Violating Background}

\author{Mariana Frank and Ismail Turan  \\
Department of Physics, Concordia University,\\
7141 Sherbrooke St. West, Montreal, Quebec, H4B 1R6, CANADA\\
E-mails: \email{mfrank@alcor.concordia.ca} and \email{ituran@physics.concordia.ca}} 
\author{\.{I}smet Yurdu\c{s}en \\
Centre de Recherches Math\'{e}matiques, Universit\'{e} de Montr\'{e}al,\\
CP 6128, Succ. Centre-Ville, Montr\'{e}al, Qu\'{e}bec H3C 3J7, CANADA\\
E-mail: \email{yurdusen@crm.umontreal.ca}}

\abstract{
We systematically analyze the integrability of a Pauli system in Lorentz violating background at the non-relativistic level both in two- and three-dimensions. We consider the non-relativistic limit of the Dirac equation from the QED sector of the so-called Standard Model Extension by keeping only two types of background couplings, the vector $a_\mu$ and the axial vector $b_\mu$. We  show that the spin-orbit interaction comes as a higher order correction in the non-relativistic limit of the Dirac equation.  Such an interaction allows  the inclusion of spin degree non-trivially, and if  Lorentz violating terms are allowed,  they might be comparable under  special circumstances. By including  all possible first-order derivative terms and considering the cases $\bm{a}\ne 0,\bm{b}\ne 0$, and $b_0\ne 0$ one at a time,  we determine the possible forms of constants of motion operator, and discuss the existence or continuity of integrability due to Lorentz violating background.}
\preprint{CUMQ/HEP 146}


\begin{document}
%  \begin{center}
%  \huge \date{\today}
% \end{center}
%%%%%%%%%%%%%%%%%%%%%%%%%%%%%%%%%%%%%%%%%%%%%%%%%%%%%%%%%%%%%%%%%%%%%%%
\section{Introduction} \label{sec:intro} 
%%%%%%%%%%%%%%%%%%%%%%%%%%%%%%%%%%%%%%%%%%%%%%%%%%%%%%%%%%%%%%%%%%%%%%%
In classical mechanics, integrability of a Hamiltonian system can be defined as the existence of a set of $n$ functionally independent constants of motion $X_1, X_2,...,X_n$ including the Hamiltonian itself if the system has $n$ degrees of freedom. These constants of motion must be in involution. The concept of integrability can be extended to so-called superintegrability which requires at least one additional constant of motion (minimally superintegrable) and allows totally $n-1$ additional constants of motion (maximally superintegrable) \footnote{This is due to the fact that the phase space is $2n$ dimensional and at least one free degree of freedom has to be left in order to have dynamics in the system.}. These additional constants of motion, however, are not necessarily in involution among each other, nor with $X_1, X_2,...,X_{n-1}$. In a very similar manner, the notion of (super)integrability can also be defined in quantum mechanics through  well-defined linear constants of motion operators which are now supposed to be algebraically independent \cite{Bargmann, Tauch, Fris, Kalnins}. In quantum mechanics, integrability not only simplifies the calculation of energy levels and wave functions, but also provides a complete set of quantum numbers, which characterize the system completely. Superintegrability, on the other hand, may entail exact solvability. The harmonic oscillator \cite{Tauch}, and the Kepler, or Coulomb system~\cite{Bargmann} are well known examples of superintegrability. 

Systematic searching for superintegrability in Hamiltonian systems with  kinetic energy quadratic in momenta and with only coordinate dependent potentials was started quite some time ago \cite{Fris}. In these works the constants of motion were considered up to second-order in momenta in order to search for quadratic (super)integrability. Both in classical and quantum mechanics, second-order integrability of such Hamiltonians mimics the separations of variables which is the first step for exact solvability. Such close connection doesn't hold for the systems with velocity dependent potentials $V(\bm{p},\bm{A})$ \cite{Dorizzi}. 

Recently, the (super)integrability of spin-dependent Hamiltonian systems with a generic scalar potential has been studied in a systematic way up to first-order \cite{ismet}.  There are other known systems which introduce velocity-dependent interactions without requiring an external vector potential $\bm{A}$. For example, if one allows violation of Lorentz symmetry at the relativistic level, the extra fields which describe such violation can also induce velocity-dependent interactions at non-relativistic level. The aim of this study is to systematically analyze (super)integrability of a Hamiltonian system (more precisely the hydrogen atom) with spin-orbit interactions in some Lorentz violating backgrounds both in 2- and 3-dimensions. 

There exists a well established framework \cite{SME}, called Standard-Model Extension (SME), for the study of the Lorentz and Charge-Parity-Time reversal ($CPT$) violation. The Standard Model Extension is the generalization of the Standard Model (SM) with additional 
 Lorentz and $CPT$ violating interactions introduced through some tensorial background fields. It can be considered a low-energy limit of a fundamental theory in which the Lorentz and $CPT$ symmetries are exact but broken spontaneously when evolved down to low energy scale, due to the existence of these tensorial background fields. SME is one of the elegant way to formulate the problem. Of course the coordinate independence of physical observables are maintained by requiring {\it observer} Lorentz invariance which describes the transformations of coordinates. Lorentz violation considered in the model is so called {\it particle} Lorentz violation which describes rotations and boosts of particles and localized fields but not background fields in a specific observer's inertial frame. Formulating the model in non-Minkowski spacetimes \cite{SMEgrav} leads to spacetime dependent coupling coefficients. Our approach here is minimalistic, i.e., constant coefficients.

In this study we consider only the effects of vectorial background couplings $a_\mu$ and $b_\mu$ to the (super)integrability of what we call the Pauli system. Here $a_\mu$ and $b_\mu$ are related to the vacuum expectation values of some vectorial background fields. Seeking (super)integrability for a Pauli system in such LV background and aiming exact solvability  could be of some interest under certain circumstances. For example, the current bound on $b_0$ \cite{b0bound}, the time component of $b_\mu$, is $b_0^{-1}\gtrsim 10^{-3}\, cm$ and it is much weaker than that on $|\bm{b}|$, the space part of the $b_\mu$. So, for a system with timelike $b_\mu$ coupling the perturbative approach fails if the effective size of the system is of the order of $10^{-3}\, cm$, and then the search for an exact solution becomes unavoidable.

The outline of the paper is as follows. In the next section, Section~\ref{sec:model}, we  introduce the relevant part of the SME Lagrangian and consider the non-relativistic limit. Then, in Section~\ref{sec:2D}, the 2-dimensional Pauli system in LV background is discussed in three separate cases. In Section~{\ref{sec:3D}, before discussing the generalization of the problem to 3-dimensions, we present the determining equations in the most general form. We then shortly discuss the perturbative approaches to the  system in Section~{\ref{sec:pert}} and conclude in Section~{\ref{sec:concl}.

%%%%%%%%%%%%%%%%%%%%%%%%%%%%%%%%%%%%%%%%%%%%%%%%%%%%%%%%%%%%%%%%%%%%%%%
\section{The model and its non-relativistic limit} \label{sec:model} 
%%%%%%%%%%%%%%%%%%%%%%%%%%%%%%%%%%%%%%%%%%%%%%%%%%%%%%%%%%%%%%%%%%%%%%%
There exist vast number of studies on the SME both at theoretical and phenomenological levels \cite{cpt04,SMEreviews}. Here we work in the QED sector of the SME for electron \cite{Kostelecky:1999zh, photonth1, photonth2}. The QED Lagrangian for electron can be given as
\begin{eqnarray}
 {\cal L}_{\rm electron}&=&\frac{i}{2}\bar{\psi}\,\Gamma^\mu\stackrel{\leftrightarrow}{D}_\mu\psi - \bar{\psi}M\psi\,,\nonumber\\
\Gamma_\mu &=& \gamma_\mu +c_{\lambda\mu}\gamma^\lambda + d_{\lambda\mu}\gamma_5\gamma^\lambda+...\nonumber\\
M &=& m + a_\mu \gamma^\mu +  b_\mu\gamma_5\gamma^\mu + \frac{1}{2}H_{\mu\nu}\sigma^{\mu\nu}\,,
\label{QEDL}
\end{eqnarray}
where $D_\mu = \partial_\mu + i e A_\mu$ with the vector potential $A_\mu$ ($f\stackrel{\leftrightarrow}{D}_\mu g\equiv fD_\mu\,g -(D_\mu f)\,g$),  $a_\mu$ and $b_\mu$ are the (pseudo-)vectorial $CPT$-odd coefficients and the other terms are $CPT$-even tensorial coefficients. These are the only terms that are obtained from the SME but $\Gamma_\mu$ can have more terms originating from non-renormalizable higher-dimensional operators. 

It is clear from the above Lagrangian that LV doesn't require $CPT$ violation but the vice versa is true \cite{Greenberg}. In this study, we consider the only two $CPT$-odd vectorial couplings $a_\mu$ and $b_\mu$ and neglect the tensorial ones\footnote{Note that allowing non-renormalizable higher-dimensional operators induces additional vectorial fields.}. A comprehensive analysis of field redefinitions 
and redundant parameters of the model can be found in Ref.~\cite{Colladay_and_McDonald}. For example, the $a_\mu$ term can be absorbed by redefining the fermion field by a phase factor $exp(ia_\mu x^\mu)$ and thus the spectrum will be unaffected by such term. Depending on the complexity of the model (photon interaction, fermion mixings, etc.), this situation could change. A similar transformation can be found for the $b_\mu$ case if, for example, a free massless fermion is assumed, but in general the $b_\mu$ term is non-trivial. So, in our discussion we keep both $a_\mu$ and $b_\mu$ terms.

We now consider the non-relativistic limit of the Dirac equation for electron starting from Eq.~(\ref{QEDL}) with LV vectorial couplings only. We take the external electromagnetic field as $A_\mu = (A_0,\,\bm{0})$. The Dirac equation becomes\footnote{We use the natural units, $\hbar=1, c=1$, throughout the paper.} 
\begin{eqnarray}
(p_\mu\gamma^\mu-eA_\mu\gamma^\mu-a_\mu\gamma^\mu-\gamma_5b_\mu\gamma^\mu-m)\psi=0\,,
\end{eqnarray}
where we adopt the Dirac representation for the $\gamma$ matrices. For convenience one can define the energy of the electron $p_0=p_0^\prime + m$ (by extracting the rest energy) so that $p_0^\prime$ becomes much smaller than $m$ in the non-relativistic limit. If we also assume $eA_0,|a|,|b|\ll m$, we obtain the following equation for the $2\times 1$ spinor $\psi_1$, which is the large component of $\psi$, 
\begin{eqnarray}
&&\left[(p_0^\prime - m - a_0 - \bm{\sigma}\cdot\bm{b})-\frac{1}{2 m}(\bm{\sigma}\cdot(\bm{p}-\bm{a})-b_0)\,\Lambda^{-1}\, (\bm{\sigma}\cdot(\bm{p}-\bm{a})-b_0)\right]\psi_1 = 0\,,\nonumber\\
&&\Lambda \equiv \left(1+\frac{p_0^\prime-eA_0-a_0-\bm{\sigma}\cdot \bm{b}}{2m}\right)\,,
\end{eqnarray}
where $\bm{\sigma}=(\sigma_1, \sigma_2, \sigma_3)$ are the Pauli matrices. Keeping the first two terms in $\Lambda^{-1}$ and  making some straightforward manipulations lead to the following equation
\begin{eqnarray}
&\displaystyle i\frac{\partial}{\partial t}\psi_1 = H_{\rm NR} \psi_1\,,\nonumber\\
&\displaystyle H_{\rm NR}= \frac{\bm{p}^2}{2m}+eA_0-\frac{e}{4m^2}\,\bm{\sigma}\cdot \bm{E}\times \bm{p}-\frac{\bm{p}^4}{8m^3}+\frac{i e}{4m^2}\,\bm{E}\cdot\bm{p}+H_{\rm LV}\,,\nonumber\\
&\displaystyle \displaystyle H_{\rm LV}=-\frac{1}{m}\,\bm{a}\cdot\bm{p}+\bm{\sigma}\cdot\bm{b}-\frac{b_0}{m}\,\bm{\sigma}\cdot\bm{p}+\frac{a_0}{4 m^2}\,\bm{p}^2+\frac{1}{4m^2}\,\bm{\sigma}\cdot\bm{p}\,\bm{\sigma}\cdot\bm{b}\,,
\label{NR}
\end{eqnarray}
where $\bm{E}$ is the electric field and we keep up to linear terms in LV parameters, since they are presumed small, and we also neglect the constant terms. In $H_{\rm NR}$, the first order correction terms to the usual Hamiltonian are the spin-orbit, relativistic, and the potential energy correction, respectively. We assume that the potential energy is spherically symmetric, $V_0(r)\equiv eA_0(r)$ so that the term $\bm{\sigma}\cdot\bm{E}\times\bm{p}$ with $\bm{E}=\frac{1}{r}\frac{dA_0}{d\,r}\,\bm{r}$ leads to the usual spin-orbit interaction $\bm{\sigma}\cdot \bm{L}$ with $\bm{L}$  the angular momentum operator.

Had we neglected all these three corrections and considered the Lorentz symmetry preserved, there would be no spin dependence in the system (since we set the external magnetic field zero) and the spin degree of the electron would become trivial. The only spin dependence is through $\bm{\sigma}\cdot \bm{L}$ and we intend to keep that term and neglect the other two whose presence would considerably complicate our study. As seen from Eq.~(\ref{NR}), $H_{\rm LV}$ also brings spin dependent interactions only through $b_\mu$. The last two terms of $H_{\rm LV}$ come from the second term in $\Lambda^{-1}$ expansion and are higher order with respect to the first three terms of $H_{\rm LV}$. Hence, we neglect them in the rest of our calculation. We further set $m=1$ for simplicity.

Taking into account all of the above remarks, and inserting the Coulomb potential of the nucleus, the Lorentz violating Hamiltonian $H_{\rm NR}$ for the hydrogen atom  follows from Eq.~(\ref{NR}) as
\begin{eqnarray}
H=\frac{1}{2} \Big(p_1^2+p_2^2+p_3^2\Big) + V_0({r}) 
+ V_1({r})\,\bm{\sigma} \!\cdot \! 
\bm{L} - \bm{a} \!\cdot \! \bm{p} + 
\bm{\sigma} \!\cdot \! \bm{b} -
b_0\,\bm{\sigma} \!\cdot\! \bm{p}
\,, \label{LVHamiltonian}
\end{eqnarray}
where $V_0=\Gamma_1/r$, $V_1=\Gamma_2/r^3$ with  $\Gamma_1=-e^2$ and $\Gamma_2=-e^2/4$. In the absence of the LV terms we call the above Hamiltonian the Pauli system, even though a system with an external magnetic field is usually considered as Pauli system. The properties of the LV terms under the Charge (C), Parity (P), Time-reversal (T), and rotational symmetries are given in Table~\ref{UCPT}

\begin{table}[htb]
\vskip -0.5cm
	\caption{Properties of LV terms in the Hamiltonian Eq.~(\ref{LVHamiltonian}) under $C$, $P$, $T$, and rotational symmetries.} \label{UCPT}     \begin{center}
    \begin{tabular}{cccccc}
   % \multicolumn{1}{|c|}{Parameter} &
   %             \multicolumn{1}{|c|}{Value}
    \hline\hline
                                      & $\rm Rotation\;$ &$\;C \;$& $P\;$&$\;T\;$ & $CPT\;$ \\ \hline 
    $\bm{a}\cdot \bm{p}$              & -      &    - &  -   &  -    & - \\
    $\bm{\sigma}\cdot \bm{b}$         & -      &    + &  +   &  -    & - \\
    $b_0\, \bm{\sigma}\cdot \bm{p}$   & +      &    + &  -   &  +    & - \\
    \hline \hline
        \end{tabular}
        \end{center}
\end{table}

In order to investigate the first-order integrability and/or superintegrability
we analyze the commutativity of a general first-order integral of motion
\begin{eqnarray}
X=(A_0+\bm{A} \cdot \bm{\sigma})p_1+(B_0+\bm{B} \cdot \bm{\sigma})p_2+
(C_0+\bm{C} \cdot \bm{\sigma})p_3+\Phi_0+\bm{\Phi}\,\cdot \bm{\sigma} 
\nonumber \\
-\frac{i}{2}\left\{(A_0+\bm{A} \cdot \bm{\sigma})_x
+(B_0+\bm{B} \cdot \bm{\sigma})_y+(C_0+\bm{C} \cdot \bm{\sigma})_z
\right\}
\,, \label{generalintofmotion}
\end{eqnarray}
with the Hamiltonian given in Eq. (\ref{LVHamiltonian}). In Eq. (\ref{generalintofmotion}), $\bm{A}$, $\bm{B}$, $\bm{C}$ and $\bm{\Phi}$ where $\bm{A}=(A_1,A_2,A_3)$ ($\bm{B}$, $\bm{C}$ and $\bm{\Phi}$ are defined similarly) are real functions of $\bm{r}$. The commutativity relation 
\begin{equation} 
[H,X]=0\,,\label{commutator}
\end{equation}
has second-, first-,  and zeroth-order terms in the momenta. Setting the coefficients of each of these terms equal to zero we get  equations determining $A_0$, $B_0$, $C_0$, $\Phi_0$ and $A_i$, $B_i$, $C_i$, $\Phi_i$ ($i=1,2,3$). 

%%%%%%%%%%%%%%%%%%%%%%%%%%%%%%%%%%%%%%%%%%%%%%%%%%%%%%%%%%%%%%%%%%%%%%%
\section{The analysis of 2D Pauli system in LV background}
\label{sec:2D}
%%%%%%%%%%%%%%%%%%%%%%%%%%%%%%%%%%%%%%%%%%%%%%%%%%%%%%%%%%%%%%%%%%%%%%%
In this section we consider the integrability of 2D Pauli system in LV background. There are three LV terms in the Hamiltonian given in Eq.~(\ref{LVHamiltonian}). For the sake of simplicity we consider the effects of each term one at a time; the case with non-zero vector coupling ($a_{i} \ne 0$), with pure spacelike axial-vector coupling ($b_i \ne 0$), and finally with pure timelike axial-vector coupling ($b_0 \ne 0$).

%%%%%%%%%%%%%%%%%%%%%%%%%%%%%%%%%%%%%%%%%%%%%%%%%%%%%%%%%%%%%%%%%%%%%%%
\subsection{Lorentz violation with vector coupling ($a_{i} \ne 0$) :}
\label{subsec:2D1}
%%%%%%%%%%%%%%%%%%%%%%%%%%%%%%%%%%%%%%%%%%%%%%%%%%%%%%%%%%%%%%%%%%%%%%%
In the absence of the axial-vector coupling terms in the Hamiltonian, Eq. (\ref{LVHamiltonian}) in two-dimensions is given as
\begin{eqnarray}
\displaystyle
H_{\rm 2D}^{(a_\mu\ne 0)} =\frac{1}{2} \Big(p_1^2+p_2^2\Big) + V_0({\rho}) 
+ V_1({\rho})\,\sigma_3 L_3 - a_1p_1-a_2p_2
\,, \label{LVHamiltonianveccoup2D}
\end{eqnarray}
where again $V_0=\Gamma_1/\rho$ and  $V_1=\Gamma_2/\rho^3$ , $\Gamma_1$ and $\Gamma_2$ are constants defined in the previous section. Since now the Hamiltonian Eq. (\ref{LVHamiltonianveccoup2D}) is a diagonal matrix  we can also choose $X$ given in Eq. (\ref{generalintofmotion}) as diagonal too and write
\begin{eqnarray}
X_{\rm 2D}^{(a_\mu\ne 0)}&=&(A_0+A_3\sigma_3)p_1+(B_0+B_3\sigma_3)p_2+
\Phi_0+\Phi_3 \sigma_3\nonumber\\
&&-\frac{i}{2}\left\{(A_0+A_3\sigma_3)_x
+(B_0+B_3\sigma_3)_y
\right\}
\,. \label{intofmotionvc2D}
\end{eqnarray}
The requirement of vanishing of the commutator $[H_{\rm 2D}^{(a_\mu\ne 0)}, X_{\rm 2D}^{(a_\mu\ne 0)}]$ gives us a total of $12$ equations for $A_{\nu}$, $B_{\nu}$ and $\Phi_{\nu}$, ($\nu=0,3$). The technique is to start from the highest order terms (second-order in our problem) and determine some functions in $X_{\rm 2D}^{(a_\mu\ne 0)}$ and then use these solutions to apply the same procedure repeatedly for the remaining orders until all unknown functions in $X_{\rm 2D}^{(a_\mu\ne 0)}$ are fixed. So, let's start from the second-order. We get the following six partial differential equations\footnote{Throughout the paper, the subscripts $x$, $y$ and $z$ represent the partial derivatives with respect to the cartesian coordinates.}
\begin{eqnarray}
A_{\nu x}=0\,, \qquad B_{\nu y}=0\,, \qquad  A_{\nu y}+B_{\nu x}=0\,,\quad (\nu=0,3)\,,  \label{vc2Dsecondorder0}
\end{eqnarray}
which can immediately be integrated to give 
\begin{eqnarray}
A_{\nu}=\xi_{\nu}+w_{\nu}y\,, \qquad   B_{\nu}=\eta_{\nu} - w_{\nu}x\,.  \label{vc2Dsecondorder}
\end{eqnarray}

After introducing Eq. (\ref{vc2Dsecondorder}) into the coefficients of the first-order terms and separating the real and imaginary parts we have $4$ determining equations for $\Phi_{\nu}$
\begin{eqnarray}
\displaystyle
\Phi_{\nu x}&=& \delta_{\nu, 3-\kappa}[(w_{\kappa}x-\eta_{\kappa})y V_{1 y}-(w_{\kappa}y+\xi_{\kappa})y V_{1 x}-\eta_{\kappa}V_1]+a_2w_{\nu}\,, \nonumber \\
\Phi_{\nu y}&=& \delta_{\nu, 3-\kappa}[(w_{\kappa}y+\xi_{\kappa})x V_{1 x}-(w_{\kappa}x-\eta_{\kappa})x V_{1 y}+\xi_{\kappa}V_1]-a_1w_{\nu}\,, \;\; (\nu=0,3)\,, \label{vc2Dfirstorder}
\end{eqnarray}
where $\delta_{\nu, \kappa}$ is the Kronecker delta function.
In order to satisfy the compatibility conditions for $\Phi_{\nu}$ we must have $\xi_{\nu}=0$ and $\eta_{\nu}=0,\;\;\nu=0,3 $. Then, Eq. (\ref{vc2Dfirstorder}) can be easily integrated to give 
\begin{eqnarray}
\Phi_{\nu}=w_{\nu}(a_2x-a_1y)\,, \qquad (\nu=0,3)\,, \label{vc2Dphis}
\end{eqnarray}
where we used $V_1=\Gamma_2/\rho^3$ and set the integration constants equal to zero without loss of generality.

Two of the $12$ determining equations, which are indeed the coefficients of the zeroth-order terms, have not been used up to now. If we introduce all the information gathered from the coefficients of the higher-order terms into the coefficients of the zeroth-order terms we get the following two conditions to be satisfied
\begin{eqnarray}
(a_1x+a_2y)V_1w_{\nu}=0\,, \qquad (\nu=0,3)\,. \label{vc2Dzeroorder}
\end{eqnarray}
Clearly we must have either (1) both $a_1=0$ and $a_2=0$, the components of the LV vector $a_\mu$ on the plane or (2) $w_{\nu}=0$. 

Case (1) corresponds to a LV vector of the form $a_\mu = (a_0,0,0,a_3)$ or even a pure timelike vector ($a_3=0$). The latter is trivial since the system still remains Pauli system up to some constants. The former is non-trivial since a non-zero space component $a_3$ is allowed but it doesn't couple with the electron in $xy$ plane. Therefore, under these conditions only, the symmetries of the Pauli system are restored. For example, we get the constant of motion as
\begin{eqnarray}
X_{\rm 2D}^{(a_\mu\ne 0)}(a_1=0=a_2)&=& -(w_0+w_3 \sigma_3)(x p_2 -y p_1)\nonumber\\
&=&-(w_0+w_3 \sigma_3) L_3\,,
\label{vc2Dconst}
\end{eqnarray}
where $w_0$ and $w_3$ are arbitrary constants. As expected both $L_3\otimes I$ and $L_3\otimes\sigma_3$ are constants of motion. However, they are not different from each other since $\sigma_3$ trivially commutes with the Hamiltonian in 2D. Therefore, there is only one extra constant of motion other than the Hamiltonian and there exists first-order integrability.

For the Case (2), which is indeed the non-trivial one, $X_{\rm 2D}^{(a_\mu\ne 0)}$ vanishes as seen from Eq.~(\ref{vc2Dconst}) and there is no first-order integrability.

%%%%%%%%%%%%%%%%%%%%%%%%%%%%%%%%%%%%%%%%%%%%%%%%%%%%%%%%%%%%%%%%%%%%%%%
\subsection{Lorentz violation with pure spacelike axial-vector coupling ($b_i \ne 0$):}
\label{subsec:2D2}
%%%%%%%%%%%%%%%%%%%%%%%%%%%%%%%%%%%%%%%%%%%%%%%%%%%%%%%%%%%%%%%%%%%%%%%
If we only have pure spacelike axial-vector coupling, i.e., $b_i \ne 0$, then in two-dimensions the Hamiltonian Eq. (\ref{LVHamiltonian}) and the integral of motion Eq. (\ref{generalintofmotion}) are given as
\begin{eqnarray}
H_{\rm 2D}^{(b_i\ne 0)}&=&\frac{1}{2} \Big(p_1^2+p_2^2\Big) + V_0({\rho}) 
+ V_1({\rho})\,\sigma_3 L_3 + \bm{\sigma} \!\cdot \! \bm{b} 
\,, \label{LVHamiltonianaxialveccoup2Dsl} \\
X_{\rm 2D}^{(b_i\ne 0)}&=&(A_0+\bm{A} \cdot \bm{\sigma})p_1+(B_0+\bm{B} \cdot \bm{\sigma})p_2+\Phi_0+\bm{\Phi}\,\cdot \bm{\sigma}\nonumber\\
&&-\frac{i}{2}\left\{(A_0+\bm{A} \cdot \bm{\sigma})_x
+(B_0+\bm{B} \cdot \bm{\sigma})_y
\right\}
\,, \label{intofmotionavc2Dsl}
\end{eqnarray}
where $V_0=\Gamma_1/\rho$ and $V_1=\Gamma_2/\rho^3$ are assumed. Since the Hamiltonian Eq. (\ref{LVHamiltonianaxialveccoup2Dsl}) is not diagonal, we can no longer choose the integral of motion operator $X_{\rm 2D}^{(b_i\ne 0)}$ diagonal.

In this case, unlike the previous case, there are $12$ determining equations coming from the coefficients of the second-order terms since we have non-diagonal entries which double the number of equations. The ones from the  diagonal elements are exactly the same as Eq. (\ref{vc2Dsecondorder0}) and hence the solutions for $A_{\nu}$ and $B_{\nu},\;(\nu=0,3)$ as given in Eq. (\ref{vc2Dsecondorder}), are the same. From the off-diagonal elements we also have the following $6$ equations
\begin{eqnarray}
&2yA_1V_1+A_{2x}=0\,, \qquad
2yA_2V_1-A_{1x}=0\,, \qquad
\nonumber \\
&2xB_2V_1+B_{1y}=0\,, \qquad
2xB_1V_1-B_{2y}=0\,, \qquad
\nonumber \\
&2V_1(xA_2-yB_2)+A_{1y}+B_{1x}=0\,, \qquad
2V_1(xA_1-yB_1)-A_{2y}+B_{2x}=0\,.  \label{avc2Dsloffdiag}
\end{eqnarray}
However, there is only a so called {\it trivial solution} (keeping in mind that  $V_1=\Gamma_2/\rho^3$) for the system of equations given in Eq. (\ref{avc2Dsloffdiag}), which is $A_1=A_2=B_1=B_2=0$. This has been shown both by hand and by using the Maple software \cite{maple}.

When we introduce all of this information into the rest of the determining equations we get, in addition to the equations given in Eq. (\ref{vc2Dfirstorder}) (with of course $a_1=0$ and $a_2=0$), the following four equations for the contribution of the coefficients of the first-order terms 
\begin{eqnarray}
&2b_2w_3y+2yV_1\Phi_2-\Phi_{1x}=0\,, \qquad 
2b_1w_3y+2yV_1\Phi_1+\Phi_{2x}=0\,, \qquad
\nonumber \\
&2b_2w_3x+2xV_1\Phi_2+\Phi_{1y}=0\,, \qquad 
2b_1w_3x+2xV_1\Phi_1-\Phi_{2y}=0\,.
\end{eqnarray}
These equations can be solved and we find that $\xi_{\nu}=0$, $\eta_{\nu}=0$, $\Phi_0=0\;(\nu=0,3)$ are solutions if either of the following conditions is satisfied
\begin{itemize}
 \item Case (1): $b_1=0=b_2$
 \item Case (2): $w_3 = 0$\,.
\end{itemize}

\noindent Note that the determining equations coming from the coefficients of the zeroth-order terms are identically satisfied in either case. The constant of motion operator is the same as given in Eq.~(\ref{vc2Dphis}) but, since this time there is no condition on $w_0$,  $L_3\otimes I$ becomes a constant of motion. It is also important to note here that since the Hamiltonian Eq.~(\ref{LVHamiltonianaxialveccoup2Dsl}) has $\sigma_1$ and $\sigma_2$ in it, the fact that $\sigma_3$ commutes with the Hamiltonian is no longer {\it trivial}. Hence, the term $L_3\otimes \sigma_3$ that comes with $w_3$ should now be considered as independent.  Obviously, in the presence of $b_1$ and $b_2$ (Case (2)) we must have $w_3=0$ which leaves $L_3\otimes I$ as the only constant of motion. However, Case (1) puts no constraint on the z-component of $b_\mu$, and if the coupling vector is perpendicular to the motion of the plane ($b_z\ne 0$), both $L_3\otimes I$ and $L_3\otimes \sigma_3$ are constants of motion but of course they are not independent. Therefore, including the Hamiltonian there are two constants of motion, independent of which of the above cases are satisfied. The first-order integrability is preserved in the presence of a LV background $\bm{b}$ with even arbitrary nonzero components in each direction.

%%%%%%%%%%%%%%%%%%%%%%%%%%%%%%%%%%%%%%%%%%%%%%%%%%%%%%%%%%%%%%%%%%%%%%%
\subsection{Lorentz violation with pure timelike axial-vector coupling ($b_0 \ne 0$):}
\label{subsec:2D3}
%%%%%%%%%%%%%%%%%%%%%%%%%%%%%%%%%%%%%%%%%%%%%%%%%%%%%%%%%%%%%%%%%%%%%%%
When $\bm{a}=\bm{0}$, $\bm{b}=\bm{0}$ but $b_0\ne 0$ the Hamiltonian Eq. (\ref{LVHamiltonian}) and the integral of motion Eq. (\ref{generalintofmotion}) reduce to the following in two-dimensions 
\begin{eqnarray}
H_{\rm 2D}^{(b_0\ne 0)}&=&\frac{1}{2} \Big(p_1^2+p_2^2\Big) + V_0({\rho}) 
+ V_1({\rho})\,\sigma_3 L_3 - b_0\big(\sigma_1p_1+\sigma_2p_2\big) 
\,, \label{LVHamiltonianaxialveccoup2Dtl} \\
X_{\rm 2D}^{(b_0\ne 0)}&=&(A_0+\bm{A} \cdot \bm{\sigma})p_1+(B_0+\bm{B} \cdot \bm{\sigma})p_2+\Phi_0+\bm{\Phi}\,\cdot \bm{\sigma} \nonumber \\
&&-\frac{i}{2}\left\{(A_0+\bm{A} \cdot \bm{\sigma})_x
+(B_0+\bm{B} \cdot \bm{\sigma})_y
\right\}
\,, \label{intofmotionavc2Dtl}
\end{eqnarray}
where $V_0=\Gamma_1/\rho$ and $V_1=\Gamma_2/\rho^3$ are assumed. Again, because of the fact that the Hamiltonian Eq. (\ref{LVHamiltonianaxialveccoup2Dtl}) is not diagonal, we can no longer choose the integral of motion diagonal as in the previous case.

In this case we have $12$ equations coming from the coefficients of the second-order terms. Three of them are exactly the same with Eq. (\ref{vc2Dsecondorder0}) for $\nu=0$ and hence we have the same $A_0$ and $B_0$ as in Eq. (\ref{vc2Dsecondorder}). The rest of the determining equations from the coefficients of the second-order terms can be written as
\begin{eqnarray}
&2b_0A_2+A_{3x}=0\,, \quad
2b_0B_1-B_{3y}=0\,, \quad
2yA_2V_1-A_{1x}=0\,, \quad
2xB_1V_1-B_{2y}=0\,, \nonumber \\
&\!\!\!\!2b_0A_3-2yA_1V_1-A_{2x}=0\,, \quad
2b_0B_3+2xB_2V_1+B_{1y}=0\,, \quad
2b_0A_1-2b_0B_2-A_{3y}-B_{3x}=0\,, \nonumber \\
&\!\!\!\!\!\!\!\!\!\!\!\!2b_0A_3+2V_1\big(xA_2-yB_2\big)+A_{1y}+B_{1x}=0\,, \quad
2b_0B_3+2V_1\big(xA_1-yB_1\big)-A_{2y}-B_{2x}=0\,.\nonumber\\ \label{avc2Dtlsecorderdet}
\end{eqnarray}
There is only a so called trivial solution for the system of equations given in Eq. (\ref{avc2Dtlsecorderdet}), which is $\bm{A}=\bm{0}$ and  $\bm{B}=\bm{0}$. This is shown again by using the Maple software \cite{maple}.

After introducing the already found functions $A_0$,  $B_0$, $\bm{A}$, and $\bm{B}$ into the determining equations coming from the coefficients of the first-order terms we have the following $8$ partial differential equations for $\Phi_0$ and $\bm{\Phi}$ 
\begin{eqnarray}
&\Phi_{0x}=0\,, \qquad
\Phi_{0y}=0\,, \nonumber \\
&\Phi_{1x}=2yV_1\Phi_2\,, \qquad
\Phi_{1y}=-2xV_1\Phi_2-b_0(w_0+2\Phi_3)\,, \nonumber \\
&\Phi_{2x}=-2yV_1\Phi_1+b_0(w_0+2\Phi_3)\,, \qquad
\Phi_{2y}=2xV_1\Phi_1\,,\nonumber \\
&\Phi_{3x}=-y(w_0y+\xi_0)V_{1x}+y(w_0x-\eta_0)V_{1y}-\eta_0V_1-2b_0\Phi_2\,, \nonumber \\
&\Phi_{3y}=x(w_0y+\xi_0)V_{1x}-x(w_0x-\eta_0)V_{1y}+\xi_0V_1+2b_0\Phi_1\,. \label{avctlfirstorder}
\end{eqnarray}
It is immediately seen that $\Phi_0=\rm{const}$, which can be taken as zero without loss of generality. Then we are left with $6$ first-order partial differential for $\bm{\Phi}$. The requirement of the equality of the mixed partial derivatives give $3$ more equations for $\bm{\Phi}$ and their first-order derivatives. Now, if we eliminate the first-order derivatives of $\bm{\Phi}$ in this system by using the equations given in Eq. (\ref{avctlfirstorder})  we get a system of algebraic equations for $\bm{\Phi}$. In order to solve this system the rank of the coefficient matrix has to be equal to the rank of the extended matrix, determined by adding the non homogeneous vector as a column. A simple analysis shows that this condition can only be possible if $\xi_0=0$ and $\eta_0=0$. In this case we have the following solution
\begin{equation}
\Phi_1=-\frac{y}{x}\Phi_2\,, \qquad \Phi_3=-\frac{w_0}{2}+\left(\frac{1}{2b_0x}+\frac{b_0}{V_{1,\,x}}\right)\Phi_2\,. \label{avc2Dfirstphis}
\end{equation}

Finally, we get the following $4$ partial differential equations from the coefficients of the zeroth-order terms 
\begin{eqnarray}
&&2b_0\Phi_{3x}-2V_1(y\Phi_{1x}-x\Phi_{1y})-(\Phi_{2xx}+\Phi_{2yy})=0\,, \quad
2b_0(\Phi_{1y}-\Phi_{2x})-(\Phi_{3xx}+\Phi_{3yy})=0\,, \nonumber \\
&&2b_0\Phi_{3y}-2V_1(y\Phi_{2x}-x\Phi_{2y})+(\Phi_{1xx}+\Phi_{1yy})=0\,, \quad
y\Phi_2+x(x\Phi_{2y}-y\Phi_{2x})=0\,,\label{avc2Dtlzerothdet}
\end{eqnarray}
where $\bm{A}=\bm{0}$, $\bm{B}=\bm{0}$, $A_0=w_0\, y$ and $B_0=-w_0\, x$ are used. Together with Eq. (\ref{avc2Dfirstphis}) it is possible to show that only $\Phi_2=0$ satisfies simultaneously all the $4$ differential equations in Eq. (\ref{avc2Dtlzerothdet}). Thus, to sum up we find the following set of solutions in the case if $b_0\ne 0$:
\begin{eqnarray}
\displaystyle
&A_0=w_0\, y\,, \quad B_0=-w_0\, x\,, \quad \bm{A}=\bm{0}\,, \quad \bm{B}=\bm{0}\,,\nonumber\\
&\Phi_{0}=0\,, \quad \Phi_{1}=0\,, \quad \Phi_{2}=0\,, \quad \Phi_{3}=-w_0/2\,. \label{avctlresult}
\end{eqnarray}
It is seen from the constant of motion operator in Eq.~(\ref{intofmotionavc2Dtl}) that we have 
\begin{eqnarray}
X_{\rm 2D}^{(b_0\ne 0)}&=&w_0(yp_1-xp_2)-\sigma_3 w_0/2\nonumber\\
&=&-w_0(L_3+\sigma_3/2)
\end{eqnarray}
which is nothing but the total angular momentum $J_3$. It is important to note here that, as opposed to the case in 2D where there are no LV terms and hence both $L_3$ and $J_3$ are constants of motion, only $J_3$ is a constant of motion in this case ( $b_0\ne 0$). The fact that $L_3$ does not commute with the Hamiltonian is something expected from the form of the LV term in Eq.~(\ref{LVHamiltonianaxialveccoup2Dtl}). However, first-order integrability is still restored.  

%%%%%%%%%%%%%%%%%%%%%%%%%%%%%%%%%%%%%%%%%%%%%%%%%%%%%%%%%%%%%%%%%%%%%%%
\section{The analysis of 3D Pauli system in LV background}
\label{sec:3D}
%%%%%%%%%%%%%%%%%%%%%%%%%%%%%%%%%%%%%%%%%%%%%%%%%%%%%%%%%%%%%%%%%%%%%%%
In general in the presence of LV terms we have the Hamiltonian and integral of motion given in equations Eq. (\ref{LVHamiltonian}) and Eq. (\ref{generalintofmotion}) respectively in $3$-dimensional Euclidean spaces. In a similar fashion as in the two-dimensional case it is also convenient to analyze the problem in three separate cases. However, it is advantageous to give the determining equations coming from the coefficients of the second- and first-order terms in full generality (that is, keeping $\bm{a},\bm{b}$, and $b_0$ non zero) before analyzing them in case by case. 

%========================================================================
\paragraph {\bf i) Determining equations coming from the second-order terms:}
%========================================================================
From the diagonal elements it is immediately found that we have the following $6$ determining equations
\begin{eqnarray}
&A_{0x}=0\,, \qquad B_{0y}=0\,, \qquad  C_{0z}=0\,,  \nonumber \\
&A_{0y}+B_{0x}=0\,, \quad A_{0z}+C_{0x}=0\,, \quad B_{0z}+C_{0y}=0\,,
\end{eqnarray}
from which the following solutions are obvious
\begin{eqnarray}
A_0=\beta_1-\alpha_3 y+\alpha_2 z \,, \nonumber\\
B_0=\beta_2+\alpha_3 x-\alpha_1 z \,, \nonumber\\
C_0=\beta_3-\alpha_2 x+\alpha_1 y \,. \label{A0B0C0}
\end{eqnarray}
where $\alpha_i$ and $\beta_i$ ($i=1,2,3$) are integration constants.  Note that these solutions are independent of the LV background couplings so that they hold for each of the cases discussed below. If we compare Eq. (\ref{A0B0C0}) with Eq. (\ref{vc2Dsecondorder}) we see that we have the following correspondences: $\beta_1=\xi_0$, $\beta_2=\eta_0$ and $\alpha_3=-w_0$. After introducing the equation (\ref{A0B0C0}) into the rest of the coefficients of the second-order terms and separating the imaginary and real parts of the coefficients coming from the off-diagonal elements we are left with an over determined system of $18$ partial differential equations for $A_i$, $B_i$, $C_i$ ($i=1,2,3$). These are,
\begin{eqnarray}
2zA_1V_1+A_{3 x}+2 b_0 A_2=0\,, \label{setof18eqs1} \\
2yA_1V_1+A_{2 x}-2 b_0 A_3=0\,, \label{setof18eqs2} \\
2xB_2V_1+B_{1 y}+2 b_0 B_3=0\,, \label{setof18eqs3} \\
2zB_2V_1+B_{3 y}-2 b_0 B_1=0\,, \label{setof18eqs4} \\
2xC_3V_1+C_{1 z}-2 b_0 C_2=0\,, \label{setof18eqs5} \\
2yC_3V_1+C_{2 z}+2 b_0 C_1=0\,, \label{setof18eqs6}
\end{eqnarray}
\begin{eqnarray}
2V_1\big(yA_2+zA_3\big)-A_{1 x}=0\,, \label{setof18eqs7} \\
2V_1\big(xB_1+zB_3\big)-B_{2 y}=0\,, \label{setof18eqs8} \\
2V_1\big(xC_1+yC_2\big)-C_{3 z}=0\,, \label{setof18eqs9} 
\end{eqnarray}
\begin{eqnarray}
2zV_1\big(A_2+B_1\big)+A_{3 y}+B_{3 x}+2 b_0(B_2-A_1)=0\,, \label{setof18eqs10} \\
2yV_1\big(A_3+C_1\big)+A_{2 z}+C_{2 x}+2 b_0(A_1-C_3)=0\,, \label{setof18eqs11} \\
2xV_1\big(B_3+C_2\big)+B_{1 z}+C_{1 y}+2 b_0(C_3-B_2)=0\,, \label{setof18eqs12}
\end{eqnarray}
\begin{eqnarray}
2V_1\big(xA_1+yA_2-zC_1\big)-A_{3 z}-C_{3 x}-2 b_0 C_2=0\,, \label{setof18eqs13} \\
2V_1\big(xB_1+yB_2-zC_2\big)-B_{3 z}-C_{3 y}+2 b_0 C_1=0\,, \label{setof18eqs14} \\
2V_1\big(xA_2-yB_2-zB_3\big)+A_{1 y}+B_{1 x}+2 b_0 A_3=0\,, \label{setof18eqs15} \\
2V_1\big(xA_1+zA_3-yB_1\big)-A_{2 y}-B_{2 x}+2 b_0 B_3=0\,, \label{setof18eqs16} \\
2V_1\big(xA_3-yC_2-zC_3\big)+A_{1 z}+C_{1 x}-2 b_0 A_2=0\,, \label{setof18eqs17} \\
2V_1\big(yB_3-xC_1-zC_3\big)+B_{2 z}+C_{2 y}+2 b_0 B_1=0\,. \label{setof18eqs18}
\end{eqnarray}

%========================================================================
\paragraph {\bf ii) Determining equations coming from the first-order terms:}
%========================================================================
We set the coefficients of the $p_i$ ($i=1,2,3 $) zero at each entry of the commutation relation. After introducing equations (\ref{A0B0C0}) and separating the real and imaginary parts, we have the following $21$ partial differential equations, 12 of which are
\begin{eqnarray}
%%%%%%%%%%%%%%%%%%%%%%%%%%%%%%%%%%%%%%%%%%%%%%%% ---1--- %%%%%%%%%%%%%%%%%%%%%%%%%%%%%%%%
&\hspace*{-1cm}x\, \bm{\hat r}\cdot\bm{\beta}\,\dot{V_1} +(\beta _1- y \alpha _3+2 y \Phi _3) V_1+\Phi _{2 z}-\bm{a}\cdot \bm{\nabla} C_2- b_0(\alpha _1-2 \Phi _1)-2(\bm{b}\times \bm{C})_2 = 0\,, \label{coeffirstord1} \\
%%%%%%%%%%%%%%%%%%%%%%%%%%%%%%%%%%%%%%%%%%%%%%%% ---2--- %%%%%%%%%%%%%%%%%%%%%%%%%%%%%%%%
&\hspace*{-1cm}x\, \bm{\hat r}\cdot\bm{\beta}\,\dot{V_1} + (\beta _1+z \alpha _2-2 z \Phi _2) V_1- \Phi _{3 y}+\bm{a}\cdot \bm{\nabla} B_3- b_0 (\alpha _1-2 \Phi _1)+2(\bm{b}\times \bm{B})_3= 0\,, \label{coeffirstord2} \\
%%%%%%%%%%%%%%%%%%%%%%%%%%%%%%%%%%%%%%%%%%%%%%%%%%% ---3--- %%%%%%%%%%%%%%%%%%%%%%%%%%%%%%%%%%%%%
&\hspace*{-1cm}y\, \bm{\hat r}\cdot\bm{\beta}\,\dot{V_1}+ (\beta _2-z \alpha _1 +2 z  \Phi _1)V_1+ \Phi _{3 x}-\bm{a}\cdot \bm{\nabla} A_3- b_0(\alpha _2-2\Phi _2)
-2(\bm{b}\times\bm{A})_3=0\,, \label{coeffirstord3} \\
%%%%%%%%%%%%%%%%%%%%%%%%%%%%%%%%%%%%%%%%%%%%%%%%%% ---4--- %%%%%%%%%%%%%%%%%%%%%%%%%%%%%%%%%%%%%
&\hspace*{-1cm}y\, \bm{\hat r}\cdot\bm{\beta}\,\dot{V_1}+ (\beta _2+x \alpha _3  -2 x \Phi _3)V_1- \Phi _{1 z}+\bm{a}\cdot \bm{\nabla} C_1-b_0(\alpha _2- 2\Phi_2)+2(\bm{b}\times\bm{C})_1=0\,, \label{coeffirstord4} \\
%%%%%%%%%%%%%%%%%%%%%%%%%%%%%%%%%%%%%%%%%%%%%%%%%% ---5--- %%%%%%%%%%%%%%%%%%%%%%%%%%%%%%%%%%%%%
&\hspace*{-1cm}z\, \bm{\hat r}\cdot\bm{\beta}\,\dot{V_1}+ (\beta _3-x \alpha _2  +2 x \Phi _2)V_1+ \Phi _{1 y} - \bm{a}\cdot \bm{\nabla} B_1 - b_0( \alpha _3 -2 \Phi _3)-2(\bm{b}\times\bm{B})_1=0\,, \label{coeffirstord5} \\
%%%%%%%%%%%%%%%%%%%%%%%%%%%%%%%%%%%%%%%%%%%%%%%%%%%%% ---6--- %%%%%%%%%%%%%%%%%%%%%%%%%%%%%%%%%%%
&\hspace*{-1cm}z\, \bm{\hat r}\cdot\bm{\beta}\,\dot{V_1}+ (\beta _3+y \alpha _1-2 y  \Phi _1)V_1- \Phi _{2 x}+ \bm{a}\cdot \bm{\nabla} A_2 - b_0 (\alpha _3-2 \Phi _3)+2(\bm{b}\times\bm{A})_2=0\,, \label{coeffirstord6} \\
%%%%%%%%%%%%%%%%%%%%%%%%%%%%%%%%%%%%%%%%%%%%%%%%%%%% ---7--- %%%%%%%%%%%%%%%%%%%%%%%%%%%%%%%%%%%%
&(y \alpha _2+ z \alpha _3 -2 y \Phi _2-2 z \Phi _3)V_1+ \Phi _{1 x}-\bm{a}\cdot \bm{\nabla} A_1-2 (\bm{b}\times\bm{A})_1=0\,, \label{coeffirstord7} \\
%%%%%%%%%%%%%%%%%%%%%%%%%%%%%%%%%%%%%%%%%%%%%%%%%%% ---8--- %%%%%%%%%%%%%%%%%%%%%%%%%%%%%%%%%%%%%%
&(x \alpha _1 + z \alpha _3-2 x \Phi _1-2 z \Phi _3)V_1 + \Phi _{2 y}-\bm{a}\cdot \bm{\nabla} B_2-2 (\bm{b}\times\bm{B})_2=0\,, \label{coeffirstord8} \\
%%%%%%%%%%%%%%%%%%%%%%%%%%%%%%%%%%%%%%%%%%%%%%%%%% ---9--- %%%%%%%%%%%%%%%%%%%%%%%%%%%%%%%%%%%%%%%
&(x \alpha _1+ y \alpha _2-2 x \Phi _1-2 y \Phi _2)V_1+ \Phi _{3 z}-\bm{a}\cdot \bm{\nabla} C_3-2 (\bm{b}\times\bm{C})_3=0\,, \label{coeffirstord9}
\end{eqnarray}
\begin{eqnarray}
\Phi_{0x}&=& \Big((yA_{3x}-xA_{3y})+(xA_{2z}-zA_{2x})+(zA_{1y}-yA_{1z})+(C_2-B_3)\Big)V_1 \nonumber \\
&\,&- \Big(x(\bm{\hat r}\times\bm{A})_1+y(\bm{\hat r}\times\bm{B})_1+z(\bm{\hat r}\times\bm{C})_1\Big) \dot{V_1}
-(\bm{a}\times\bm{\alpha})_1+ b_0 \bm{\nabla}\cdot\bm{A}\,, \label{coeffirstordgen10} \\
\Phi_{0y}&=& \Big((yB_{3x}-xB_{3y})+(xB_{2z}-zB_{2x})+(zB_{1y}-yB_{1z})+(A_3-C_1)\Big)V_1 \nonumber \\
&\,&- \Big(x(\bm{\hat r}\times\bm{A})_2+y(\bm{\hat r}\times\bm{B})_2+z(\bm{\hat r}\times\bm{C})_2\Big) \dot{V_1}
-(\bm{a}\times\bm{\alpha})_2+ b_0 \bm{\nabla}\cdot\bm{B}\,, \label{coeffirstordgen11} \\
\Phi_{0z}&=& \Big((yC_{3x}-xC_{3y})+(xC_{2z}-zC_{2x})+(zC_{1y}-yC_{1z})+(B_1-A_2)\Big)V_1 \nonumber \\
&\,& -  \Big(x(\bm{\hat r}\times\bm{A})_3+y(\bm{\hat r}\times\bm{B})_3+z(\bm{\hat r}\times\bm{C})_3\Big) \dot{V_1}
-(\bm{a}\times\bm{\alpha})_3+ b_0 \bm{\nabla}\cdot\bm{C}\,, \label{coeffirstordgen12}
\end{eqnarray}
where $\dot{V_1}\equiv dV_1/dr$, $\bm{\hat r}$ is the unit displacement vector, $\bm{a}=(a_1,a_2,a_3)$ and $\bm{b}=(b_1,b_2,b_3)$ are the space parts of the Lorentz violating parameters $a_\mu$ and $b_\mu$ respectively, and $\bm{A}=(A_1,A_2,A_3)$ ($\bm{B}$ and $\bm{C}$ are defined similarly) is introduced in the integral of motion $X$. The rest of the determining equations from the coefficients of the first-order terms can be expressed in terms of certain derivative combinations of the determining equations from the coefficients of the second-order terms given in Eqs.~(\ref{setof18eqs1}) - (\ref{setof18eqs18}). For completeness they are symbolically listed below:
\begin{eqnarray}
&2({\rm Eq}(\ref{setof18eqs7}))_x-({\rm Eq}(\ref{setof18eqs15}))_y-({\rm Eq}(\ref{setof18eqs17}))_z=0\,, \label{coeffirstordgen1} \\
&2({\rm Eq}(\ref{setof18eqs2}))_x-({\rm Eq}(\ref{setof18eqs16}))_y+({\rm Eq}(\ref{setof18eqs11}))_z=0\,, \label{coeffirstordgen2} \\
&2({\rm Eq}(\ref{setof18eqs1}))_x+({\rm Eq}(\ref{setof18eqs10}))_y-({\rm Eq}(\ref{setof18eqs13}))_z=0\,, \label{coeffirstordgen3} \\
&2({\rm Eq}(\ref{setof18eqs3}))_y+({\rm Eq}(\ref{setof18eqs15}))_x+({\rm Eq}(\ref{setof18eqs12}))_z=0\,, \label{coeffirstordgen4} \\
&2({\rm Eq}(\ref{setof18eqs8}))_y+({\rm Eq}(\ref{setof18eqs16}))_x-({\rm Eq}(\ref{setof18eqs18}))_z=0\,, \label{coeffirstordgen5} \\
&2({\rm Eq}(\ref{setof18eqs4}))_y+({\rm Eq}(\ref{setof18eqs10}))_x-({\rm Eq}(\ref{setof18eqs14}))_z=0\,, \label{coeffirstordgen6} \\
&2({\rm Eq}(\ref{setof18eqs5}))_z+({\rm Eq}(\ref{setof18eqs17}))_x+({\rm Eq}(\ref{setof18eqs12}))_y=0\,, \label{coeffirstordgen7} \\
&2({\rm Eq}(\ref{setof18eqs6}))_z+({\rm Eq}(\ref{setof18eqs11}))_x+({\rm Eq}(\ref{setof18eqs18}))_y=0\,, \label{coeffirstordgen8} \\
&2({\rm Eq}(\ref{setof18eqs9}))_z+({\rm Eq}(\ref{setof18eqs13}))_x+({\rm Eq}(\ref{setof18eqs14}))_y=0\,, \label{coeffirstordgen9}
\end{eqnarray}
where $(...)_x$ for example represents the derivative of the entire equation with respect to $x$.

 We do not include the expressions for the 8 determining differential equations coming from the zeroth-order terms. They are very lengthy and their exact form is not particularly illuminating. We simplify the equations  by imposing the solutions obtained from the second- and first-order terms, and then they are either trivially satisfied or reduce to simple constraint equations (like Eq.~(\ref{vc2Dzeroorder}) in the 2D case), presented for each case in the following discussion.

In all three cases to be discussed below we have the same integral of motion Eq.~(\ref{generalintofmotion}) and  the functions $A_0$, $B_0$ and $C_0$ are obtained the same as in Eq.~(\ref{A0B0C0}). Therefore, we will briefly discuss the solutions for the rest of the coefficients by applying the technique used so far.
 
%%%%%%%%%%%%%%%%%%%%%%%%%%%%%%%%%%%%%%%%%%%%%%%%%%%%%%%%%%%%%%%%%%%%%%%
\subsection{Lorentz violation with vector coupling ($a_{i} \ne 0$) :}
\label{subsec:3D1}
%%%%%%%%%%%%%%%%%%%%%%%%%%%%%%%%%%%%%%%%%%%%%%%%%%%%%%%%%%%%%%%%%%%%%%%
In this case we have the Hamiltonian Eq.~(\ref{LVHamiltonian}) with $\bm{b}=\bm{0}$ and $b_0=0$. It is found by using Maple that when $b_0=0$ we have the following solution for the $18$ determining equations given in Eq.~(\ref{setof18eqs1}) - Eq.~(\ref{setof18eqs18})
\begin{eqnarray}
A_1=0, \qquad A_2=zw, \qquad A_3=-yw, \nonumber \\
B_1=-zw, \qquad B_2=0, \qquad B_3=xw, \nonumber \\
C_1=yw, \qquad C_2=-xw, \qquad C_3=0, \label{solutionof18eqs}
\end{eqnarray}
where $w$ is an integration constant.

After introducing Eq.~(\ref{solutionof18eqs}) into the rest of the determining equations it is immediately found from Eq.~(\ref{coeffirstordgen10}) - Eq.~(\ref{coeffirstordgen12})  that 
\begin{equation}
\Phi_0=(\alpha_2a_3-\alpha_3a_2)x+(\alpha_3a_1-\alpha_1a_3)y+(\alpha_1a_2-\alpha_2a_1)z\,, \label{3Dcase1phi0}
\end{equation}
and then we are left with $9$ first-order partial differential equations for $\bm{\Phi}=(\Phi_1,\Phi_2,\Phi_3)$. These are the equations Eq.~(\ref{coeffirstord1}) - Eq.~(\ref{coeffirstord9}) with $\bm{b}=\bm{0}$ and $b_0=0$. In order to solve this system we express the first-derivatives of $\bm{\Phi}$ and look for the compatibility of the mixed partial derivatives. The requirement of the equality of the mixed partial derivatives gives another $9$ equations for $\bm{\Phi}$  and its first-order derivatives. Now, introducing the first-derivatives of $\bm{\Phi}$ which are found from Eq.~(\ref{coeffirstord1}) - Eq.~(\ref{coeffirstord9}) into this system, we get a system of algebraic equations for $\bm{\Phi}$. In order to have a solution of this algebraic system we must have $\bm{\beta}=\bm{0}$ and $w=0$. It is not suprising to require $w=0$ since it is the coefficient of the $\bm{\sigma}\,\cdot \bm{L}$ term in Eq.~(\ref{generalintofmotion}) and clearly it cannot commute with the Hamiltonian Eq. (\ref{LVHamiltonian}) in the presence of $\bm{a}$ terms. Setting $\bm{\beta}=\bm{0}$ and $w=0$ we find  
\begin{equation}
\bm{\Phi}=\frac{\bm{\alpha}}{2}\,. 
\label{3Dcase1phis}
\end{equation}
% \begin{eqnarray}
% \Phi_1=\frac{\alpha_1}{2}\,, \qquad \Phi_2=\frac{\alpha_2}{2}\,, \qquad \Phi_3=\frac{\alpha_3}{2}\,. \label{3Dcase1phis}
% \end{eqnarray}

Finally, if we introduce all the information gathered from the coefficients of the higher-order terms into the determining equations coming from the coefficients of the zeroth-order terms, we find that the following $3$ conditions must also be satisfied
\begin{eqnarray}
a_1(y\alpha_2+z\alpha_3)-(a_2y+a_3z)\alpha_1=0\,, \nonumber \\
a_2(x\alpha_1+z\alpha_3)-(a_1x+a_3z)\alpha_2=0\,, \nonumber \\
a_3(x\alpha_1+y\alpha_2)-(a_1x+a_2y)\alpha_3=0\,. \label{3Dcase1zerothdet}
\end{eqnarray}
Clearly it is seen that in order to satisfy these $3$ conditions we must either have  $\bm{a}=\bm{0}$ or  $\bm{\alpha}=\bm{0}$. Since the former case implies that there are no LV terms and the latter one indicates we have no nontrivial first-order integral of motion,  we conclude that in the presence of $\bm{a}$ terms there is no first-order integrability unlike the 2D case, if $\bm{a}$ is perpendicular to the plane of the motion ($L_3$ as discussed is constant of motion). It should be noted that $a_\mu$ term can be eliminated with a transformation and in general is not observable physically. This is only true for simple one fermion models or non-interacting multi-fermion cases but does not hold for a generic multi-fermion theory.

%%%%%%%%%%%%%%%%%%%%%%%%%%%%%%%%%%%%%%%%%%%%%%%%%%%%%%%%%%%%%%%%%%%%%%%
\subsection{Lorentz violation with pure spacelike axial-vector coupling ($b_i \ne 0$):}
\label{subsec:3D2}
%%%%%%%%%%%%%%%%%%%%%%%%%%%%%%%%%%%%%%%%%%%%%%%%%%%%%%%%%%%%%%%%%%%%%%%
After setting $\bm{a}=\bm{0}$ and $b_0=0$ in Eq.~(\ref{LVHamiltonian}), we again find the same solution as in Eq.~(\ref{solutionof18eqs}) for $\bm{A},\bm{B},$ and $\bm{C}$ since the $18$ equations are identical in the absence of $b_0$. However, from Eq.~(\ref{coeffirstordgen10}) - Eq.~(\ref{coeffirstordgen12}), in this case we obtain $\Phi_0=\rm{const}$, which can be taken as zero without loss of generality. Then, we proceed in a similar fashion as in the previous case and reach the same conclusion for the system of $9$ differential equations given in Eq.~(\ref{coeffirstord1}) - Eq.~(\ref{coeffirstord9}), with $\bm{a}=\bm{0}$ and $b_0=0$. That is, $\bm{\Phi}=\bm{\alpha}/2$ is still the solution together with the requirement that $\bm{\beta}=\bm{0}$ and $w=0$. After introducing all the information gathered from the coefficients of the higher-order terms into the determining equations coming from the coefficients of the zeroth-order terms, we find, as in the previous case, that the following condition must also be satisfied
\begin{equation}
\bm{b}\, \cdot \bm{\alpha}=\bm{0}\,. 
\label{3Dcase2zerothdet}
\end{equation}

Again, we can conclude that in general in the presence of $\bm{b}$ terms there is no first-order integrability. However, we may find some integral of motions in some special cases (e.g. choosing $b_1=0$, $b_2=0$ and $\alpha_1=0$, $\alpha_2=0$). In general there is a constant of motion $X^{(b_i\ne 0)}_{3D}=\bm{\alpha}\cdot\bm{J}$ if $\bm{b}$ is a vector perpendicular to $\bm{\alpha}$ ( here $\bm{J}$ represents the total angular momentum operator). Of course, not all components of $\bm{\alpha}$ are independent due to Eq.~(\ref{3Dcase2zerothdet}). For example, if we assume the $i^{th}$ component of $\bm{b}$ to be nonzero, then one can alternatively write the constant of motion as 
\begin{equation}
\displaystyle
X^{(b_i\ne 0)}_{3D}=\frac{1}{b_i}\left[(\bm{\alpha\times(b\times J})\right]_i\,.
\label{3DC3CM}
\end{equation}
where $\bm{\alpha}$ is otherwise arbitrary. From here one could think that there are two constants of motion after eliminating one of the components of $\bm{\alpha}$ but this is not true. Once the orientation of $\bm{b}$ is fixed, there is a unique $\bm{\alpha}$ so that we have one additional constant of motion, which could be chosen as $J_3$. One noticeable difference from the 2D case is that the total angular momentum $J_3$ is now conserved, but this is not enough to make the system first-order integrable.

%%%%%%%%%%%%%%%%%%%%%%%%%%%%%%%%%%%%%%%%%%%%%%%%%%%%%%%%%%%%%%%%%%%%%%%
\subsection{Lorentz violation with pure timelike axial-vector coupling ($b_0 \ne 0$):}
\label{subsec:3D3}
%%%%%%%%%%%%%%%%%%%%%%%%%%%%%%%%%%%%%%%%%%%%%%%%%%%%%%%%%%%%%%%%%%%%%%%
In this case, the set of 18 equations in Eqs.~(\ref{setof18eqs1}) - (\ref{setof18eqs18}) differs from the previous $2$ cases since $b_0 \ne 0$. The only solution of this system is the so called {\it trivial solution} which is $\bm{A}=\bm{0}$, $\bm{B}=\bm{0}$ and $\bm{C}=\bm{0}$. We find also that $\Phi_0=\rm{const}$, which can be taken as zero without loss of generality, and proceed in the same way as in the previous cases. We again find the same $\bm{\Phi}$ as in Eq. (\ref{3Dcase1phis}) together with the only requirement $\bm{\beta}=\bm{0}$. However, this time the determining equations coming from the coefficients of the zeroth-order terms are identically satisfied if we introduce all the information gathered from the coefficients of the higher-order terms. The form of $X^{(b_0\ne 0)}_{3D}$ is the same as in the previous case but without the constraint Eq.~(\ref{3Dcase2zerothdet}). There exist however two additional constants of motion since the third one is related to the first two through an $SU(2)$ algebra.

Thus, we conclude that the Hamiltonian Eq. (\ref{LVHamiltonian}) with $\bm{a}=\bm{0}$ and $\bm{b}=\bm{0}$ is first-order integrable even though it has LV terms ($b_0 \ne 0$). It is easily seen that the components of $\bm{J}$ commute with $H$, hence the system is integrable. We may choose $\{J_i, \bm{J}^2, H\}$ as a commuting set of operators. It is important to note that in the commuting set we choose $\bm{J}^2$ but there is no way to get it in $X^{(b_0\ne 0)}_{3D}$ since $\bm{J}^2$ is a second-order operator. However, first-order integrals of motion $J_i$ generate $\bm{J}^2$.
 
%%%%%%%%%%%%%%%%%%%%%%%%%%%%%%%%%%%%%%%%%%%%%%%%%%%%%%%%%%%%%%%%%%%%%%%
\section{Perturbative Solutions}\label{sec:pert}
%%%%%%%%%%%%%%%%%%%%%%%%%%%%%%%%%%%%%%%%%%%%%%%%%%%%%%%%%%%%%%%%%%%%%%%
The whole idea in this study is to seek, with a systematic method, the first-order integrability/superintegrability of the Pauli system in (axial-)vectorial LV background. The ultimate goal is to achieve at least a quasi-exact solution. This could be important for the $b_0\ne 0$ case since the bound on $b_0$ for electron is much weaker than both $\bm{a}$ and $\bm{b}$, and if the effective size of the system comparable with $b_0^{-1}$, the perturbative approach fails. So, it would be enough to mainly concentrate on the $b_0\ne 0$ case and even perform a second-order analysis. However, our experience is that it would be much harder to solve the equations of overdetermined systems analytically, especially if one keeps the spin-orbit term. 

Based on our analysis we don't have first-order superintegrability in either 2D or 3D. There are first-order integrable cases in 2D under special arrangements but this is only true in 3D for the $b_0\ne 0$ case, where exact solutions could be of interest. Of course, neither integrability nor superintegrability does guarantee the separability and existence of exact solution but either is an important step in finding one. Recently, a perturbative approach has been carried out for a similar Hamiltonian system in  the same LV background within the QED version of Standard Model Extension \cite{Kharlanov:2007yp,Ferreira:2006kg}. For completeness, we like to briefly summarize their results. Note that the analysis in Ref.~\cite{Kharlanov:2007yp} includes an external magnetic field but not the spin-orbit term. A more extensive perturbative analysis for all types of Lorentz violating couplings is found in \cite{Bluhm:1997qb}.

For the case with $a_\mu$, as discussed in Section \ref{subsec:3D1}, using a field redefinion by a factor of $exp(i\, a_\mu x^\mu)$, the energy is shifted by a constant $a_0$ and wavefunction has an overall phase factor $exp(i\,\bm{a}\cdot\bm{r})$, compared to a hydrogen atom  without LV background.

For the case $\bm{b}\ne 0$, the  perturbative approach is applied \cite{Kharlanov:2007yp} and energy shifts are obtained, by an amount proportional to $b_3 m_s$ (to first order). Here $m_s$ is the magnetic spin quantum number. The wave function is unchanged. The energy correction is proportional to $b_3\, m_j/(l+1)$ if the spin-orbit correction is included. Here $m_j$ and $l$  are the total magnetic angular momentum  and the orbital angular momentum numbers, respectively. These corrections are quite small due to experimental bounds on $|\bm{b}|$.

The case with $b_0\ne 0$ is indeed interesting. It is shown that using a unitarity transformation one could relate a real hydrogen with all corrections (spin-orbit, relativistic, and the so-called Darwin term) to the one with additional LV background of the form $b_0\, \bm{\sigma\cdot p}$. The unitarity transformation is $U=1-i\,b_0(1+\Sigma/r)\bm{\sigma\cdot r}$, up to first order in $b_0$, so that $H^{(b_0\ne 0)}_{\rm real\, Hy}=U^\dagger H_{\rm real\, Hy}U$ and $\Psi^{(b_0\ne 0)}=U^\dagger \Psi$. Here $\Sigma$ is a constant appropriately chosen to obtain the correction terms. It is then straightforward to solve the system $H_{\rm real\, Hy}$ by applying perturbation theory and to transform everything back to the original system. As usual there is no contribution to the energy spectrum of the hydrogen atom from the $b_0$ term, up to first order in $b_0$.  However, the wave function develops a $b_0$ dependent part. See the Ref.~\cite{Kharlanov:2007yp} for the details.

%%%%%%%%%%%%%%%%%%%%%%%%%%%%%%%%%%%%%%%%%%%%%%%%%%%%%%%%%%%%%%%%%%%%%%%
\section{Conclusion}\label{sec:concl}
%%%%%%%%%%%%%%%%%%%%%%%%%%%%%%%%%%%%%%%%%%%%%%%%%%%%%%%%%%%%%%%%%%%%%%%
In this study we have analyzed in a systematic way the integrability of a Pauli system in a LV background both in two- and three-dimensions. We consider the Hamiltonian which originates from the non-relativistic limit of the QED part of the so-called Standard Model Extension. We kept only two types of LV couplings to the electron, the vector type of coupling represented by $a_\mu$ and the axial-vector type of coupling by $b_\mu$. Since $a_0$ enters as a constant term in the Hamiltonian, we consider in both 2D and 3D the following three separate cases; LV from $\bm{a}\ne 0$, LV from $\bm{b}\ne 0$, and LV from $b_0\ne 0$. The last case is especially important since perturbative approaches could fail under certain circumstances, due to weak experimental bounds on $b_0$, and exact solutions might be required. Note that in our study we kept the spin-orbit correction to the hydrogen atom,  partially because the correction from the LV term with $b_0\ne 0$ could be comparable with the spin-orbit term, as shown in Section \ref{sec:model}. 

After writing the most general constant of motion operator $X$ in first-order derivatives, we obtained sets of overdetermined differential equations from the vanishing of the commutator of $X$ with the Hamiltonian. In 2D, first-order integrability is possible if $\bm{a}$ is perpendicular to the plane of the motion, when the LV is due to $\bm{a}\ne 0$ only. The constant of motion is one of the components of the angular momentum operator $\bm{L}$ ($L_3$). Our conclusion is the same for $\bm{b}\ne 0$ case, with the difference that the vector $\bm{b}$ can have arbitrary components in each direction. The last case in 2D is $b_0\ne 0$ and  first-order integrability is restored like in the other cases with one difference: the $z$-component of the total angular momentum operator $J_3$ is conserved if the motion is in $xy$ plane. 

Based on our 2D results, we extended our discussion into a more realistic 3D picture by following the same procedure. The first-order integrability disappears for the $\bm{a}\ne 0$ case in 3D. For $\bm{b}\ne 0$ case, even though $J_3$ becomes a constant of motion, it is not enough to make the system integrable (since we need another one, unlike 2D case). The first-order integrability in 3D is retained only for the $b_0\ne 0$ case where we can find two additional constant of motion operators like $J_1$ and $J_2$. We believe that studying the  $b_0\ne 0$ case up to second-order is worthwhile in search of superintegrability. This is in fact the case where exact solutions could be physically relevant. One caveat is that, at the second order, solving the overdetermined partial differential equations becomes much more challenging. For completeness, we also summarized the results of a recent paper \cite{Kharlanov:2007yp} where unitary transformations and perturbative approaches are employed for similar systems.

%%%%%%%%%%%%%%%%%%%%%%%%%%%%%%%%%%%%%%%%%%%%%%%%%%%%%%%%%%%%%%%%%%%%%%%
\acknowledgments
%%%%%%%%%%%%%%%%%%%%%%%%%%%%%%%%%%%%%%%%%%%%%%%%%%%%%%%%%%%%%%%%%%%%%%%
We thank Alan Kostelecky and Pavel Winternitz for communications and helpful discussions. The work of M.F. and I.T. is supported in part by NSERC of Canada under the Grant No. SAP01105354. \.{I}.Y. acknowledges a postdoctoral fellowship awarded 
by the Laboratory of Mathematical Physics of the CRM, Universit\'{e} de Montr\'{e}al.

%%%%%%%%%%%%%%%%%%%%%%%%%%%%%%%%%%%%%%%%%%%%%%%%%%%%%%%%%%%%%%%%%%%%%%%

\end{document}